\documentclass[twocolumn,aps,prl,superscriptaddress, longbibliography]{revtex4-2}
\usepackage{mathtools}
\usepackage{bm}
\usepackage{amsmath}
\usepackage{amssymb}
\usepackage{graphicx}
\usepackage{tabularx}

\makeatletter
\usepackage{dcolumn}
\usepackage{bm}
\usepackage{multirow}

\usepackage{multirow}

\usepackage{bm}

\usepackage{color}
{\catcode `\@=11 \global\let\AddToReset=\@addtoreset}
\AddToReset{equation}{section}

\newcounter{mnotecount}[section]

\usepackage{chngcntr}
\counterwithout{equation}{section}

\makeatother

\begin{document}

\title{Linear cooling of a levitated micromagnetic cylinder by vibration}

\author{Chris Timberlake}
\email{C.J.Timberlake@soton.ac.uk}
\affiliation{School of Physics and Astronomy, University of Southampton, Southampton
SO17 1BJ, UK}

\author{Elliot Simcox}
\email{es8g18@soton.ac.uk}
\affiliation{School of Physics and Astronomy, University of Southampton, Southampton
SO17 1BJ, UK}

\author{Hendrik Ulbricht}
\email{H.Ulbricht@soton.ac.uk}
\affiliation{School of Physics and Astronomy, University of Southampton, Southampton
SO17 1BJ, UK}

\date{\today}
\begin{abstract}

We report feedback cooling of translational and librational degrees of freedom of a levitated micromagnet cylinder, utilizing a piezoelectric actuator to apply linear feedback to high-Q mechanical modes. The normal modes are measured with a superconducting pick-up coil coupled to a DC SQUID, and phase information is fed back to the piezoelectric actuator to feedback cool a center-of-mass mode to $\sim 7$~K, and a librational mode to $830 \pm 200$~mK. Q-factors of $1.0 \times 10^7$ are evaluated for the center-of-mass mode. We find that it is plausible to achieve ground state cooling of the center-of-mass mode by introducing vibration isolation, optimizing the geometry of the pick-up coil to focus on the specific mode of interest and utilizing a state-of-the-art SQUID for detection.
\end{abstract}
\maketitle

\section{Introduction}

Cooling the center-of-mass motion of macroscopic objects to their quantum ground state has long been a goal within the physics community as it is regarded as a crucial first step towards not only observing quantum mechanical effects on the macroscale - for example by generating spatial quantum superpositions of single trapped large-mass particles, aka matterwave interferometry~\cite{Arndt2014, Romero-Isart2011, Rahman2019, millen2020quantum} - but also for searching for new physics in the form of deviations from known interactions and by checking postulates of new particles~\cite{Geraci2015, Diehl2018, Monteiro2020, Fadeev2021, Moore2021}. The study of gravitational effects of massive particles in quantum states is of much interest~\cite{aspelmeyer2022zeh, ulbricht2021testing} as it might be a way to shine light on the interplay between quantum mechanics and gravity via experiments. It is understood that a larger macrosopicity of quantum states~\cite{nimmrichter2013macroscopicity} can be achieved by decoupling mechanical oscillators from their environment by different ways of levitation. Trapping and cooling the motion of large (larger than $\mu m$ length scale) particles to the quantum ground state is extremely challenging. Optical trapping techniques are suited to trapping sub-micron sized particles, and linear feedback techniques have already been utilized in levitated optomechanics to cool nanoparticles to their motional ground state~\cite{magrini2021real, Tebbenjohanns2021}. Recently, simultaneous ground state cooling of two mechanical modes was achieved~\cite{Piotrowski2023} and even the motion of the large LIGO mirrors have been cooled close to the quantum ground state by feedback~\cite{whittle2021approaching}, besides many clamped mechanical systems~\cite{aspelmeyer2014cavity}. However, the absorption and recoil of photons from the trapping field act as a dissipation limit which scales with the sixth power of the radius of the trapped particle~\cite{Jain2016}, and there is a hard decoherence limit for quantum states in optical levitation by interactions with black body and trapping laser radiation~\cite{Bateman2014}. 

Trapping of charged particles in ion traps~\cite{Millen2015, Alda2016, Bykov2019, Bullier2020, Bykov2019} offers more flexibility on particle radius, but the active electric fields required for trapping inherently induces noise, ultimately limiting the center-of-mass motional quantum states due to anticipated charge-induced decoherence effects~\cite{sonnentag2007measurement}. While ground state cooling might be very possible, it has not been achieved for particles beyond atomic ions. In general, the ability to manipulate and control the motion of trapped particles by external fields inherently comes with the introduction of noise and decoherence.

Instead, magnetically levitated oscillators, in particular Meissner levitated ones, have the potential to not only trap and cool the mechanical motion of macroscopic objects to the quantum ground state, but due to the trapping mechanisms being entirely {\it passive}, it offers the possibility of extended periods of coherent state evolution for subsequent preparation of motional quantum states~\cite{Vinante2022}. Furthermore, trapping magnets with a size range from micrometer~\cite{Vinante2020, Wang2019, Gieseler2020} to millimeter sized magnets~\cite{Timberlake2019, Raut2021, fuchs2024} and beyond~\cite{Prat-Camps2017} is possible in contrast to nanoscale trapping in optical and electrical systems. The passive trapping can be regarded as a disadvantage when it comes to controlling and manipulating the system, but we show in this paper that by modulating vibrations, which affect all forms of particle traps, one can selectively act on the center-of-mass motion of large-mass systems and cool all the way to the quantum regime.  

A further advantage of magnetic traps using superconductivity is that all dominant decoherence effects are dramatically reduced in the cryogenic environment~\cite{gonzalez2021levitodynamics}. Several schemes for ground state cooling of super-micron sized objects have been proposed, for levitated magnets~\cite{Streltsov2021, Kani2022, Fuwa2023}, levitated superconductors~\cite{Romero-Isart2012, Cirio2012}, culminating in using ground state cooled superconducting spheres for matterwave interferometry~\cite{Pino2018}. However, cooling near the ground state has yet to be achieved experimentally in levitated systems, despite efforts toward this in cooling levitated superconducting spheres~\cite{Hofer2023}. A challenge of feedback cooling macroscopic particles to the quantum ground state is that the zero-point motion $X_{\text{ZPM}} \sim \sqrt{\hbar/2m\omega_0}$ decreases with increasing mass $m$ - albeit frequency tends to somewhat decrease with increasing mass at a lesser rate. The ability to cool a particle depends on the strength of the control fields that can be exerted on the particle, which typically scales as $g = \eta X_{\text{ZPM}}$, where $\eta$ is the coupling strength to the particle's position. This scaling makes cooling to the ground state harder as mass increases~\cite{Kani2022}. Additionally, macroscopic magnets tend to levitate with low mechanical frequencies, where vibrational noise tends to dominate. Similar experiments in dry dilution refrigerators suffer from extreme vibrations due to the pulse tube that is required to cool to cryogenic temperatures~\cite{Hofer2023, fuchs2024}. Here we show that only moderate vibration isolation (factor 100 in amplitude) is required to reach the thermal noise limit in our system - which is extremely important for ground state cooling capabilities.

In this article, we demonstrate linear feedback cooling of both translational and librational modes of a levitated micromagnet in a Meissner trap by modulating vibrations with a piezoelectric actuator, and analyse in detail, by considering leading effects, that cooling to the quantum ground state with the very same technique is within scope in the near future. 

\section{Methods}

\subsection{Description of Experimental Setup}

The magnetic trap consists of a superconducting well, made of the type-I superconductor lead, with a flat elliptical base (long axis = 5~mm, short axis = 3~mm), with a lead lid attached to fully contain the magnet inside the superconductor and shield from external magnetic fields. A neodymium (NdFeB) cylindrical ferromagnetic particle (100~$\mu$m length x 200~$\mu$m diameter from SM Magnetics) is placed at the bottom of the trap, and the setup is evacuated and cooled inside a helium-3 sorption refrigerator to $\approx$ 410~mK in temperature. All pressure measurements in this letter have been corrected for helium gas pressure, and the cryogenic temperatures (see Appendix A). To dampen the effects of external vibrations the apparatus is mounted on an 850~kg granite block, which is suspended on optical table supports with compressed air (Newport S-2000 Pneumatic Vibration Isolators). When the superconducting transition temperature is reached, the particle is lifted off of the lead surface and levitates due to the Meissner currents induced in the superconductor. The walls of the well provide lateral stability while the base of the trap, combined with gravity, provide vertical confinement. The particle acts as a harmonic oscillator with three translational and two librational modes. 

\begin{figure}[t!]
    \centering
    \includegraphics[width=\linewidth]{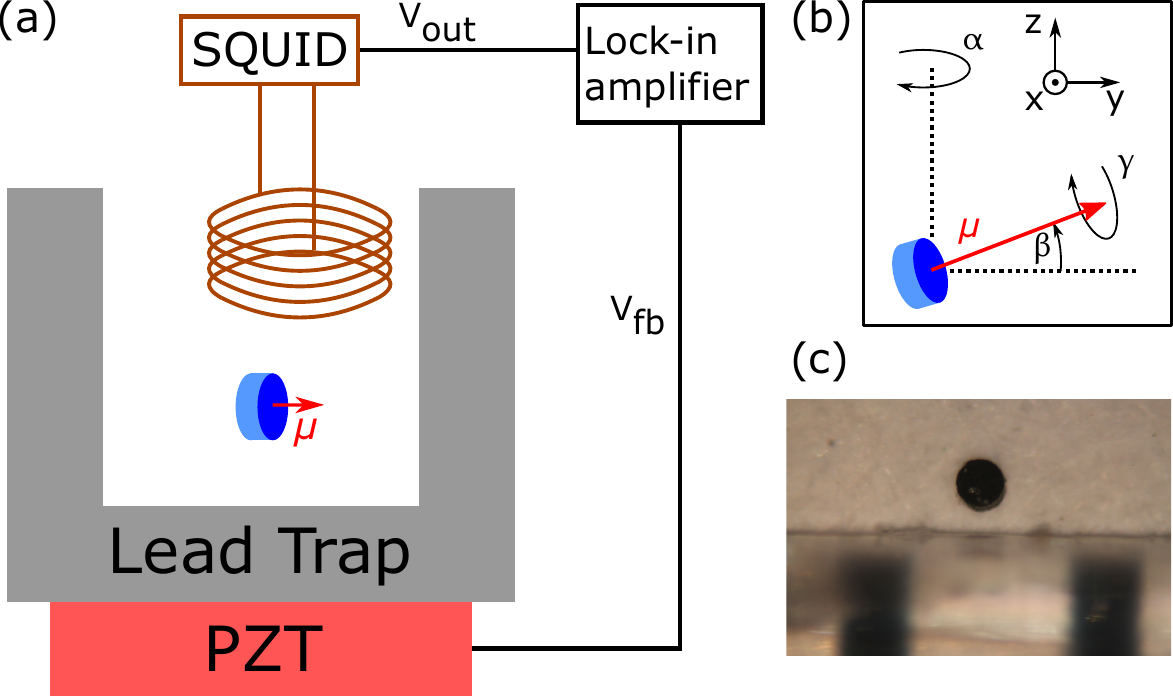}
    \caption{(a) Schematic of the experimental setup. A cylinder magnet is levitated in a lead superconducting trap. The position is measured using a pick-up coil and a SQUID, with the position information fed back to a piezoelectric actuator for cold damping. (b) The coordinate system for describing the normal modes. (c) A photograph of the NdFeB cylinder magnet, as viewed from the circular face. The magnet has a diameter of 200~$\mu$m and a thickness of 100~$\mu$m. A plastic rule is shown in the shot with 1~mm spacing for reference.}
    \label{fig:schematic}
\end{figure}

These oscillations induce a change in magnetic flux $\Phi$ in the pick-up coil, due to the magnetic dipole $\mu$ oscillating. The pick-up coil, of inductance $L$, is connected to the input coil of the SQUID (Magnicon Single Stage Current Sensor). Therefore the oscillating magnet produces a flux $\Phi_S = MI = \frac{M}{L}\Phi$ in the SQUID, where $I$ is the induced current in the pick-up coil and $M$ is the mutual inductance between the SQUID and the input coil. For these experiments the pick-up coil consists of 15 loops of 75~$\mu$m diameter NbTi wire, wrapped around a polyether ether ketone (PEEK) holder of radius of 1~mm. This pick-up coil holder is located above the magnet, inside the magnetic trap. For electromagnetic shielding purposes, the NbTi wires connecting the SQUID to the pick-up coil are in twisted pairs and fed through a lead superconducting sleeve, and the SQUID itself is housed in a Nb can.

The entire lead trap and is housed within a cryoperm shield. The motion of the particle is detected with a pick-up coil which is inductively coupled to a DC SQUID. The SQUID signal output is connected to a lock-in amplifier (Zurich Instruments HF2LI 50 MHz Lock-in Amplifier) and an oscilloscope (PicoScope 4262 Oscilloscope). The entire trap is fitted onto a piezo a piezoelectric actuator, which is used to modulate the particle motion in the vertical direction. The piezo (Thorlabs PK44LA2P2 Piezo Ring Stack) has a dynamic range of 9~$\mu$m at room temperature, and we drive it with approximated $<$0.1\% of the total 150~V voltage range. At this voltage level, we find no system heating. The amplitudes Each mode is tracked using a lock-in amplifier, which tracks the phase of the mechanical mode, which can be fed back to the piezoelectric actuator at a phase to oppose the motion, resulting in linear feedback cooling. A schematic of the setup can be seen in Fig.~\ref{fig:schematic}.

\subsection{Q-factor}

\begin{figure}[h!]
    \centering
    \includegraphics[width=\linewidth]{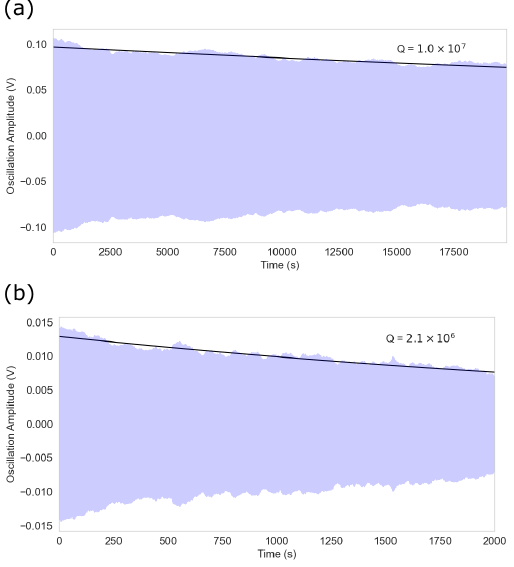}
    \caption{Ring-down measurements to evaluate the Q-factor of the $z$ (a) and $\beta$ (b) mechanical modes.}
    \label{fig:Q-factor}
\end{figure}

An important parameter used to characterise the mechanical modes of our magnet is the quality factor, or Q-factor. This Q-factor is evaluated by resonantly exciting the normal modes with the piezoelectric actuator, with a signal proportional to the negative velocity ($\pi$ phase difference to feedback cooling), and recording how the amplitude decays once the excitation field is no longer present, in a ring-down measurement. The Q-factor of the $z$ and $\beta$ modes can be seen in Fig.~\ref{fig:Q-factor} (a) and (b) respectively.

The Q-factors which are listed in the manuscript are evaluated via ringdown measurement, as shown here. The error on the Q-factor fit is determined by taking the square root on the variance, as calculated in the covariance matrix. The error for the $z$ mode ($Q_z = 1.0 \times 10^7$) is 2000, and for the beta mode ($Q_{\beta} = 2.1 \times 10^6$), the error is 2000, both to one significant figure.

\section{Results}
\subsection{Feedback cooling of damped driven oscillators}

The equation of motion for a damped, driven harmonic oscillator with an applied feedback force can be written as
\begin{align}
    \ddot{x}(t) + \Gamma_0\dot{x}(t) + \omega_0^2x(t) = \frac{F_{\text{th}}(t)+F_{\text{FB}}(t)}{m},
    \label{eq:Eq_of_motion}
\end{align}

where where $\Gamma_0$ is the background damping, as measured without feedback cooling on, $F_{\text{th}}(t)$ is the driving force due to thermal stochastic noise and $F_{\text{FB}}$ is the feedback force. The power spectral density (PSD) of a single oscillator mode, in thermal equilibrium with a thermal bath $T$, with an applied feedback force with dissipation rate $\Gamma_{\text{FB}}$ can be written as
\begin{equation}
     S_{x} (\omega) = \frac{4 k_B T \Gamma_0}{m} \frac{1}{(\omega_0^2 - \omega^2)^2 + \omega^2 (\Gamma_0 + \Gamma_{\text{FB}})^2},
    \label{eq:PSD}
\end{equation}  

where $k_B$ is the Boltzmann constant, $m$ is the mass of the oscillator, $\omega_0$ is the resonance frequency. The thermal force noise of a mechanical oscillator is given by

\begin{equation}
    S_F = 4k_BTm\omega_0/Q,
    \label{eq:thermal_noise}
\end{equation}

where $Q$ is the quality factor, defined as as $Q=\omega_0/\Gamma_0$. An analogous torque noise can be expressed as $ S_\tau = 4k_BTI\omega_0/Q,$ where $I$ is the moment of inertia. 

\begin{figure*}[t!]
    \centering
    \includegraphics[width=0.95\textwidth]{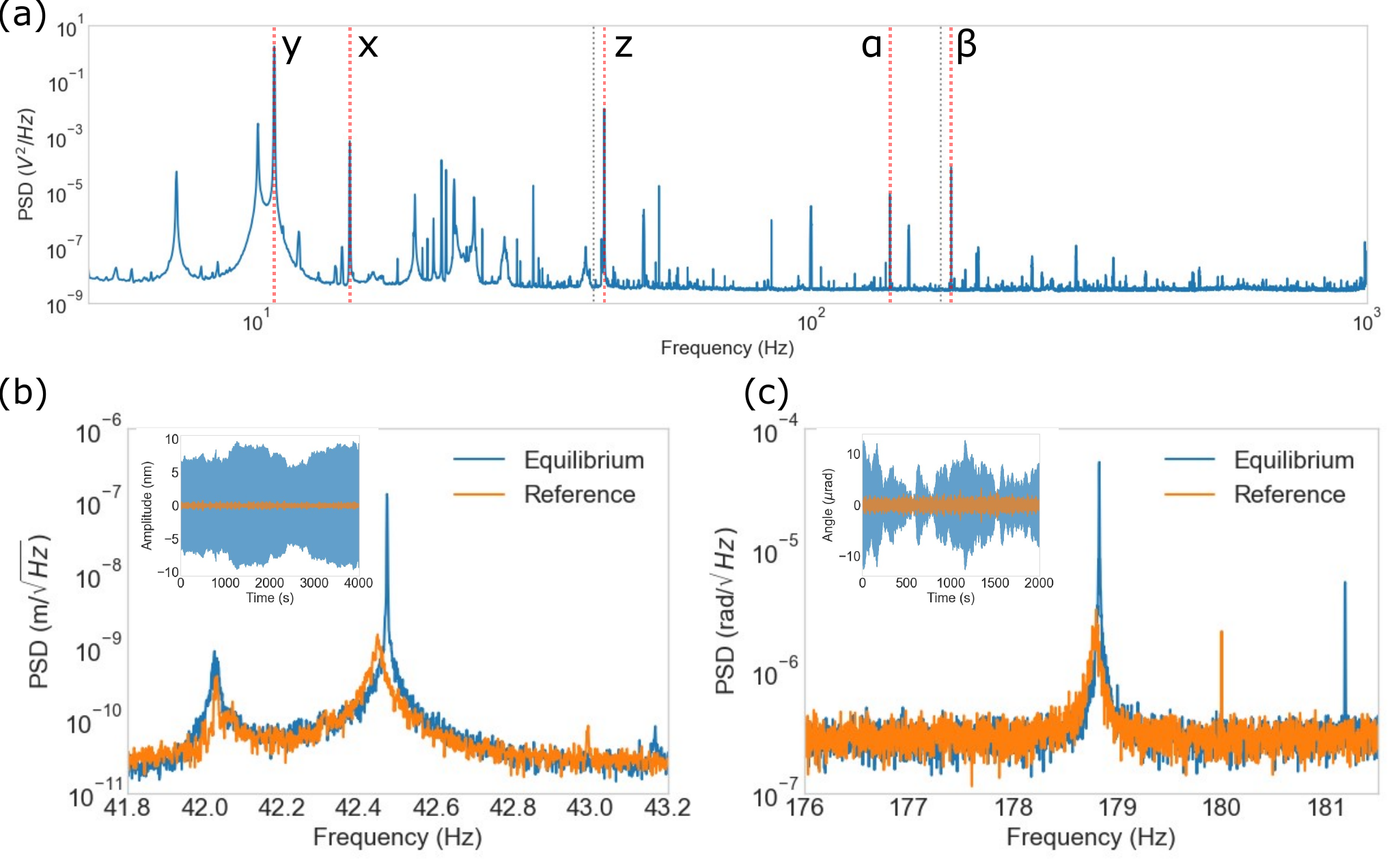}
    \caption{(a) An example spectrum of the levitated magnet, showing three translational ($x$, $y$ and $z$) and two librational modes ($\alpha$ and $\beta$). The distribution of the normal modes is identified with finite element analysis as detailed in~\cite{Vinante2020}, and we find that the $z$ and $\beta$ modes are the ones sensitive to piezo actuation. We also estimate the $z$ and $\beta$ frequencies analytically (marked as the dotted gray line) in Appendix C. (b) The power spectral density (PSD) of the $z$ mode. The orange represents the mode at a temperature of $T$ = 4.4~K and a pressure of $P$ = $2 \times 10^{-1}$~mbar, which is when the resonator is thermal noise limited and acts as a calibration (see Appendix D). The blue represents the same mode at $T$ = 410~mK and $P = 1 \times 10^{-8}$~mbar. In these conditions the magnet is no longer thermal noise limited. The inset shows the amplitude of the modes. (c) Shows the same as (b) but for the $\beta$ mode. This save was taken at $P = 4 \times 10^{-8}$~mbar. At this higher frequency, less vibrational noise is coupled into the system which results in the motion of the mode being closer to the thermal noise limit. There is a small frequency shift between reference (orange) and equilibrium (blue), of unknown origin. This frequency shift is over hours/days, and is not amplitude dependent.}
    \label{fig:spectrum}
\end{figure*}

\subsection{Motion of $z$ and $\beta$ modes}

Fig~\ref{fig:spectrum}~(a) shows the spectrum of the levitated magnet, with three translational and two librational modes identifed. We identify the distribution of the normal modes with finite element analysis as detailed in~\cite{Vinante2020}, and find that the $z$ and $\beta$ modes are the ones sensitive to piezo actuation. Figs.~\ref{fig:spectrum}~(b) and (c) show the reference PSD compared to the equilibrium data taken at $T$ = 410~mK and low pressure for the $z_{\text{eq}}$ and $\beta_{\text{eq}}$ mode respectively. The reference data is taken with $T$ = 4.4~K, which corresponds to an amplitude of $z$ = 270~pm and $\beta$ = 1.1~$\mu$rad respectively. The voltage to displacement conversion factor for the $z$ mode is given by $C_{z} = 1.76 \times 10^6$~V/m, and the analogous angle conversion factor for the $\beta$ mode is given by $C_{\beta} = 123.2$~V/rad. An equilibrium effective temperature of the $T_{\text{eq}}^{z}$ = 3400~K ($z$ = 7.6~nm) and $T_{\text{eq}}^{\beta}$ = 97~K ($\beta$ = 5.2~$\mu$rad) is evaluated. Given the geometry of our detection system (see Appendix E) relative to these amplitudes, any non-linearity in the detection is negligible ($\sim$ 0.002 \% of non-linear deviation across the amplitude range of $z$ = 7.6~nm). Considering the lack of vibration isolation, these effective temperatures are remarkably low; similar levitated systems, with comparable mass and frequency, have observed effective temperatures of $\sim10^{10}$~K in dry dilution refrigerators~\cite{Hofer2023} without any isolation ($\sim2.5$~K with isolation), and temperatures of $\sim3$~K~\cite{fuchs2024} with extensive multi-stage vibration isolation. Here, we are less than a factor of 100 in amplitude above the thermal noise floor, meaning relatively moderate vibration isolation could be implemented in our system. By undertaking ring-down measurements, we find $Q_z = 1.0 \times 10^7$, and $Q_{\beta} = 2.1 \times 10^6$. The limits of $Q$ are expected to be due to magnetic hysteresis or eddy current damping within the metallic ferromagnets, as detailed in~\cite{Prat-Camps2017}. 

By using eq.~\eqref{eq:thermal_noise} and the values measured for the $z$ mode, we can infer a thermal force noise of $S_F^{1/2} = 3.4 \times 10^{-16}$~N/$\sqrt{\text{Hz}}$ at $T_{\text{eq}}^{z}$ = 3400~K. By introducing sufficient vibration isolation, such that the mechanical motion is in thermal equilibrium with the 410~mK bath, force noises of $S_F^{1/2} = 3.7 \times 10^{-18}$~N/$\sqrt{\text{Hz}}$ could be achieved. For the $\beta$ mode, we estimate a torque noise of $S_{\tau}^{1/2} = 1.5 \times 10^{-20}$~Nm/$\sqrt{\text{Hz}}$ at $T_{\text{eq}}^{\beta}$ = 97~K, which with adequate vibrational damping could be lowered to $S_{\tau}^{1/2} = 9.8 \times 10^{-22}$~Nm/$\sqrt{\text{Hz}}$ at 410~mK. 

\begin{figure}[h!]
    \centering
    \includegraphics[width=0.9\linewidth]{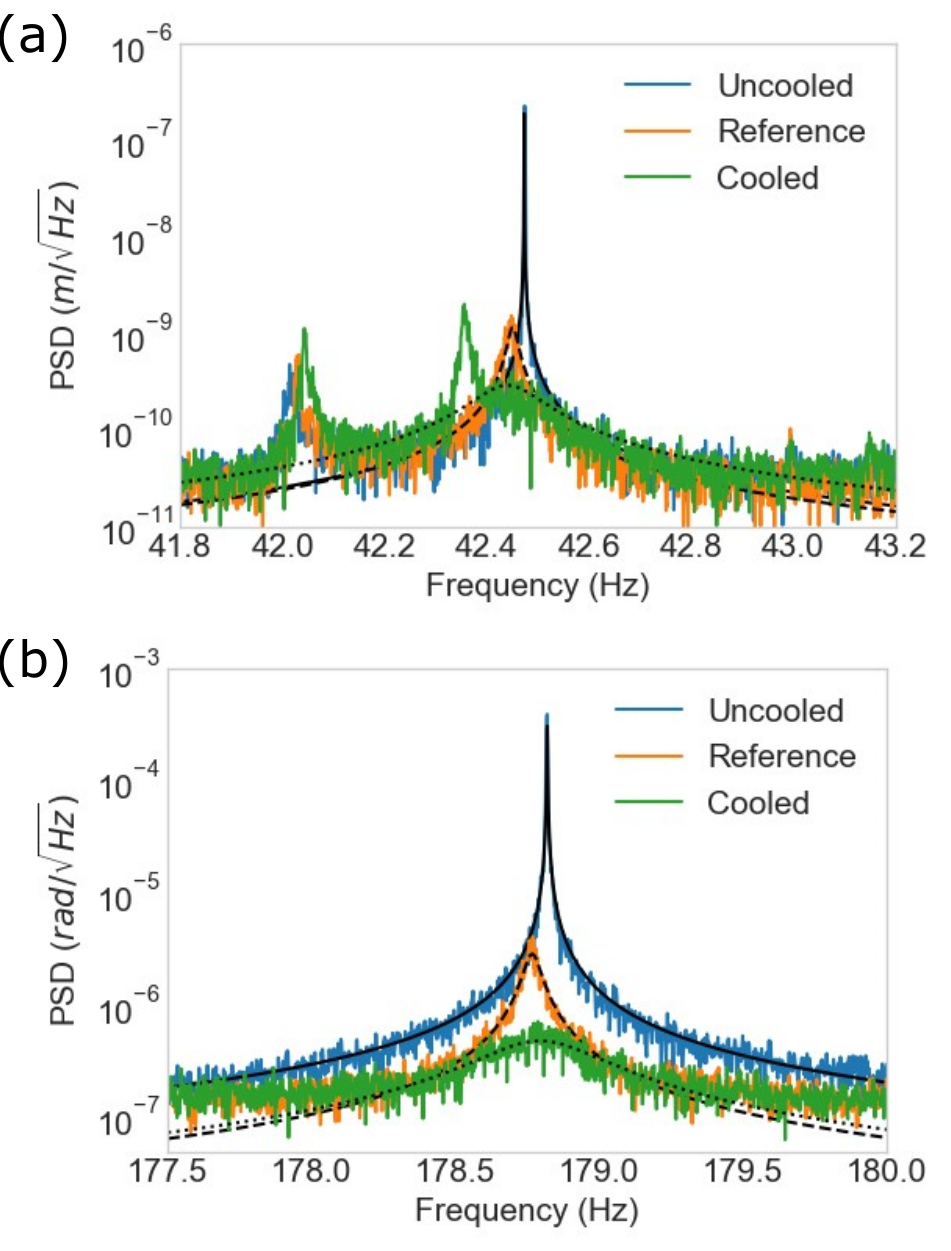}
    \caption{PSDs comparing the reference data to uncooled data and feedback cooled data. (a) Shows the 42.4~Hz translational mode. The reference save was taken at $T = 4.4$~K and $P = 2\times 10^{-1}$~mbar. Here the feedback cooled data was taken at $P = 6\times 10^{-7}$~mbar, with $T_{\text{FB}}^{z} \sim 7$~K. This temperature is extracted by comparing the RMS amplitude around the frequency peak to the reference data RMS amplitude. This is in order to account for the extra peak close to the resonance. (b) Shows the 178.8~Hz librational mode. A temperature of $T_{\text{FB}}^{\beta}$ = 830$\pm$200~mK is reached at $P = 2\times 10^{-7}$~mbar. Here the reference save was taken at $T = 4.2$~K and $P = 1\times 10^{-1}$~mbar.}
    \label{fig:PSD_cooling}
\end{figure}

\subsection{Feedback cooling of levitated micromagnet}

By fitting to the PSD of our cooled mode, we can compare to our reference PSD to calculate the cooled effective temperature. Our reference save for the center-of-mass mode was taken at a pressure of $P$ = $2 \times 10^{-1}$~mbar, where the magnet was well thermalized with the background environment of $T$ = 4.4~K (see Appendix D). For the librational mode, the reference save was taken at $T$ = 4.2~K and $P = 1\times 10^{-1}$~mbar. Assuming equipartition theorem we can write the effective temperature of the particle mode as

\begin{equation}
    T_{\text{FB}} = T\frac{\Gamma_0}{\Gamma_0 + \Gamma_{\text{FB}}},
    \label{eq:temp}
\end{equation}

More details of this derivation can be found in~\cite{Vovrosh2017, Gieseler2012}. In order to apply a feedback force, the mode of interest is frequency filtered using a lock-in amplifier, and the phase tracked with a phase-locked loop. This signal is appropriately phase shifted, such that the feedback signal is applied as a direct force proportional to the particle's velocity, and fed to a piezoelectric actuator, increasing the effective damping on the motion. The voltage gain is manually adjusted such that the cooling is maximized. This sort of feedback cooling is referred to in the literature as cold damping, velocity damping or linear feedback cooling~\cite{Cohadon1999, Poggio2007, Tebbenjohanns2019}. For the $z$ mode, an effective temperature of $T_{\text{FB}}^{z} \sim 7$~K (amplitude $z_{\text{FB}} \sim 340$~pm) is reached at $P = 2\times 10^{-7}$~mbar. The extra peak close to the resonance in Fig.~\ref{fig:PSD_cooling}~(a) is an artifact of the phase locked loop losing track of the motion, and is accounted for in the final temperature estimate. For $\beta$, $T_{\text{FB}}^{\beta} = 830 \pm 200$~mK ($\beta_{\text{FB}} = 480 \pm 100$~nrad) was evaluated at $P = 6\times 10^{-7}$~mbar (see Fig.~\ref{fig:PSD_cooling}).

\subsection{Limits of feedback cooling}

A natural question after achieving these temperatures is what is the limit of feedback cooling that could be achieved with this style of detection and feedback? One can write the minimum achievable temperature $T_{\text{min}}$ from feedback cooling, in terms of minimum phonon number $N_{\text{min}}$ as~\cite{Vinante2012, Poggio2007}

\begin{equation}
    N_{\text{min}} = \frac{k_BT_{\text{min}}}{\hbar \omega_0} = \frac{1}{2\hbar}\sqrt{S_FS_{x_{d}}},
    \label{eq:min_temp}
\end{equation}

where $S_{x_{d}}$ is the detector displacement noise and $\hbar$ is Planck's constant. In terms of minimum temperature, we have $T_{\text{min}} = \frac{{\omega_0}}{2k_B}\sqrt{S_FS_{x_{d}}}$ ($T_{\text{min}} = \frac{{\omega_0}}{2k_B}\sqrt{S_{\tau}S_{\theta_{d}}}$ for librational modes, where $S_{\theta_{d}}$ is the angular detector noise). In these experiments, we have a detector noise of $S_{x_{d}}^{1/2} = 2.1 \times 10^{-11}$~m$/\sqrt{\text{Hz}}$ for the $z$ mode and $S_{\theta_{d}}^{1/2} = 3.0 \times 10^{-7}$~rad$/\sqrt{\text{Hz}}$ for the $\beta$ mode. By reducing the vibrational noise to the below the thermal noise limit, and by measuring to the SQUID noise floor, a minimum possible temperature achieved by feedback cooling is predicted for both the $z$ and $\beta$-mode of $T_{\text{min}}^z\sim$ 5~$\mu$K ($N_{\text{min}}^z \sim 2500$) and $T_{\text{min}}^{\beta}\sim$ 80~$\mu$K ($N_{\text{min}}^{\beta} \sim 9400$) respectively. A more thorough description of experimental parameters can be seen in Appendix F.

\subsection{Optimization of SQUID coupling}

In order to further reduce the minimum achievable temperature, or phonon number, either the thermal noise of the oscillator or the detector noise must be reduced. Thermal noise reduction could be achieved by reducing the temperature further, with a dilution refrigerator. However, given the extensive costs involved in purchasing a new system, plus the extensive vibrations introduced by having a pulse tube in dry systems, it isn't a practical solution. By tailoring the geometry of the pick-up coil to the mode of interest, we can maximize the magnetic coupling to increase our detection sensitivity, as considered in Appendix E. We find that rotating the pick-up coil by 90 degrees would result in a detection noise of $S_{x_{d}}^{1/2} = 1.3 \times 10^{-16}$~m/$\sqrt{\text{Hz}}$, which translates to a minimum phonon number $N_{\text{min}}^z \sim 2$, or 5~nK in temperature (see eq.~\eqref{eq:min_temp}). Furthermore, reducing the magnet to pick-up coil distance from 2.5~mm to 2.0~mm would result in a magnetomechanical coupling which is strong enough to reach the quantum ground state.

In order to reach the ground state, the force noise must be dominated by SQUID backaction, and the detector itself be quantum limited, such that the energy resolution of the SQUID $\epsilon \approx \hbar$~\cite{Braginsky1995, Clerk2010, Caves1982}. The backaction from SQUIDs which are not quantum limited, would set a limit on cooling such that $N_{\text{min}}>1$. SQUIDs approaching such a the quantum limit have been achieved~\cite{Wakai1988, Awschalom1988}. With state-of-the-art SQUID detection, it is feasible to achieve the necessary conditions to cool towards the quantum ground state.

\section{Discussion}

In conclusion, we have demonstrated feedback cooling, utilizing cold damping with a piezoelectric actuator, of two normal modes of a levitated micromagnetic cylinder within a superconducting trap. By driving the piezo with an AC voltage on resonance, proportional to the velocity of the resonator, we have reduced the effective temperature of one center-of-mass (CoM) mode to $\sim 7$~K, while a librational mode was cooled to $830 \pm 200$~mK. Such effective temperatures have yet to be reached in the literature for levitated magnets. Our equilibrium temperatures of 3400~K and 97~K for the $z$ and $\beta$ mode respectively are within a factor of 100 of the thermal noise floor, in amplitude. This is remarkably low compared to similar levitated systems, with comparable mass and frequency. By introducing vibrational isolation, and measuring to the SQUID noise floor, we predict temperatures of $T_{\text{min}}^z \sim$ 5~$\mu$K ($N_{\text{min}}^z \sim$~2500) and $T_{\text{min}}^{\beta} \sim$ 80~$\mu$K ($N_{\text{min}}^{\beta} \sim$~9400) could be obtained with feedback cooling. By reorienting the detection pick-up coil to maximize magnetic coupling in the $z$-direction, it will be possible to have couplings strong enough to cool to the quantum ground state. Such an achievement would open up a toolbox for generating macroscopic quantum states of motion, probing fundamental physics questions and for precision sensing applications. Our Meissner-levitation scheme has exceptionally low noise and low decoherence features, giving rise to the hope of generation of marcoscopic quantum superposition states of single trapped ferromagnets in the near future~\cite{millen2020quantum}. 

\section{Acknowledgements}

We would like to thank M. Paternostro for helpful comments on the manuscript, as well as P. Connell, J. Chalk and D. Grimsey for machining the lead traps and piezoelectric actuator mount used in this study. We acknowledge support from the QuantERA grant LEMAQUME, funded by the QuantERA II ERA-NET Cofund in Quantum Technologies implemented within the EU Horizon 2020 Programme, and funding from the UK funding agency EPSRC grants (EP/W007444/1, EP/V035975/1, EP/V000624/1, EP/X009491/1), the Leverhulme Trust project {\it MONDMag} (RPG-2022-57), the EU Horizon 2020 FET-Open project {\it TeQ} (766900), the EU Horizon Europe EIC Pathfinder project {\it QuCoM} (10032223). All data supporting this study are openly available from the University of Southampton repository at https://doi.org/10.5258/SOTON/D2806.

\section*{Appendix A: Correcting Pressure Measurements}

The inner vacuum chamber (IVC) of our cryostat has a cold side, where experiments are undertaken, and a warm side, where pumps and pressure gauges are attached. All pressure measurements are taken at the warm side of the vacuum chamber with a Bayard-Alpert Pirani vacuum gauge. Such gauges measure nitrogen standard pressure, meaning a correction factor, $C$, is needed to account for the helium gas, such that the true pressure is $P = C \times P_{N_2}$. In the Pirani range ($>2 \times 10^{-2}$~mbar) a correction factor of $C = 0.8$ is used~\cite{Pirani}, whereas in the Bayard-Alpert range ($<10^{-3}$~mbar), $C = 5.9$~\cite{FRG-730}. Additionally, the pressure at the warm side of the vacuum chamber $P_w$, at a temperature $T_w,$ will differ to the pressure at the cold side of the chamber $P_c$ at temperature $T_c$ according to the Weber-Schmidt model~\cite{Roberts1956, Vinante2020}, which states that

\begin{equation}
    \frac{P_c}{P_w} = \left(\frac{T_c}{T_w}\right)^{1/2}.
\end{equation}

All listed pressure measurements in the manuscript have been corrected for helium gas and temperature.

\section*{Appendix B: Magnetic Trap Piezo Mounting}

\begin{figure}[t!]
    \centering
    \includegraphics[width=\linewidth]{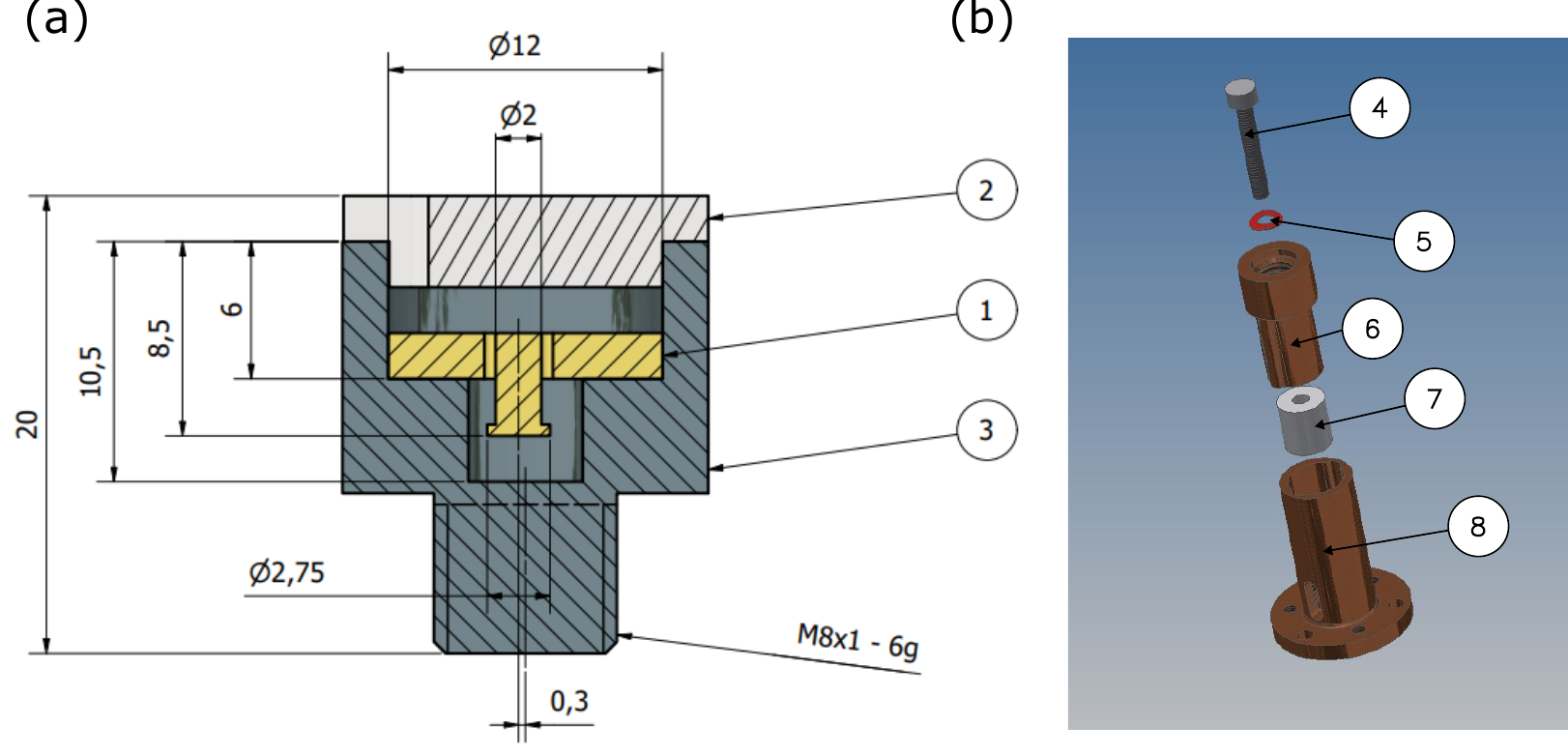}
    \caption{(a) Technical drawing of the lead magnetic trap used for the levitation experiments. The main body of the trap is represented in dark gray (3), with the PEEK pick-up coil holder in yellow (2) and the lead lid in light gray (1). The pick-up coil holder is positioned 0.3~mm off axis to break symmetry, and allow detection of five mechanical modes. (b) Exploded view of the trap mounting system with the piezoelectric actuator. (4) is the bolt which secures the spring (5), copper mount (6) and piezo (7) to the main copper support (8). The trap shown in (a) has a thread which screws into the top of (6).}
    \label{fig:Trap}
\end{figure}

For the cooling experiments, the entire trap, pick-up coil holder and trap lid are fitted onto a piezo a piezoelectric actuator, which is used to modulate the particle motion in the vertical direction. A technical drawing of the trap can be seen in Fig.~\ref{fig:Trap}(a). This is achieved by mounting the trap in a copper housing, which is then attached with a bolt, through a spring loaded piezo ring stack, as shown in Fig.~\ref{fig:Trap}~(b). 

\section*{Appendix C: $z$ and $\beta$ mode frequencies}

For our cooling experiments presented here, it's important to know which normal mode we are actuating with the piezo at any given time. From previous publications~\cite{Vinante2020}, where finite element analysis was used, we know the distribution of normal modes in our system. $y$ and $x$ are lowest, followed by $z$, then $\alpha$ and $\beta$ respectively. To distinguish the normal modes from other noise peaks, we look see which fundamental peaks respond to mechanical kicks (by lightly tapping our experiment), and also check to ensure that the linewidth of the peak narrows as the pressure inside the vacuum chamber decreases. Experimentally, we confirm that the $z$ and $\beta$ modes are where they expected by testing how they respond to small piezoelectric driving in the $z$ direction. The $z$ and $\beta$ mode are extremely responsive, whereas the $x$, $y$ and $\alpha$ mode do not get excited. 

The $z$ and $\beta$ frequencies can also be estimated analytically. According to the method of images~\cite{Jackson1998, Lin2006, Vinante2020}, the potential energy of a permanent magnet, with magnetic moment $\bm{\mu}$ and mass $m$ above a horizontal infinite superconducting plane is given by 

\begin{figure*}[t!]
    \centering
    \includegraphics[width=0.9\textwidth]{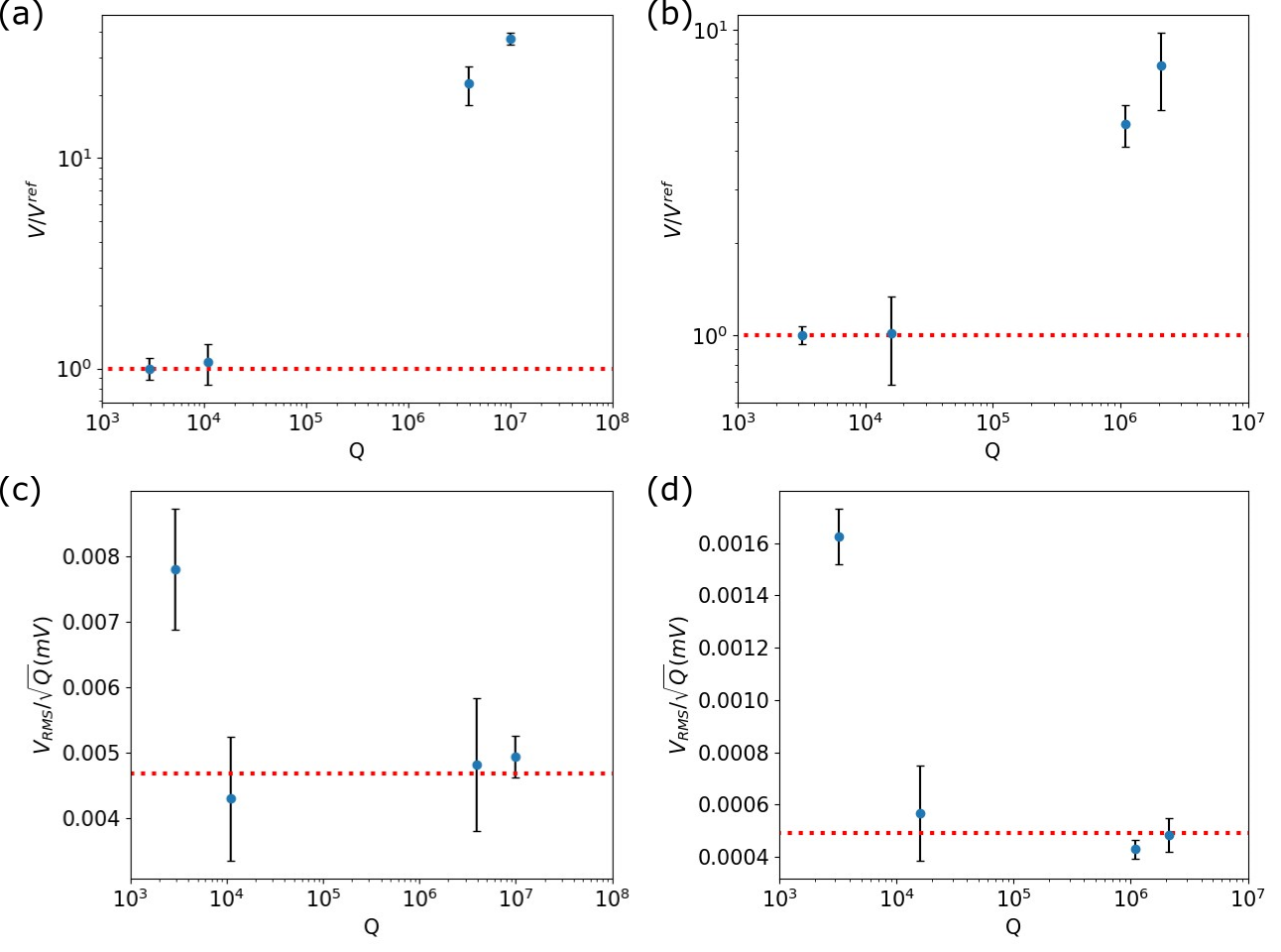}
    \caption{(a) A plot of the normalized RMS amplitude, with respect to the reference amplitude, against $Q$ for the $z$ mode. The red dotted line represents the reference save amplitude. It can be seen that the amplitude is consistent for low $Q$, up to $Q \sim 10^4$. This finding is what is expected for a thermally limited system. Above this $Q$, the thermal noise is now lower than the vibrational noise, which dominates the motion at higher $Q$. (b) Is the same as (a) but for the $\beta$ mode. (c) A plot of the ratio $V_{RMS}/\sqrt{Q}$ vs $Q$ for the $z$ mode. The ratio $V_{RMS}/\sqrt{Q}$ represents the noise driving the system, and for a thermal noise limited system should decrease as the $Q$, while remaining flat for a vibrationally limited system. We indeed see that at low $Q$, the system behaves as a thermal noise limited system is expected, and at higher $Q$ the noise power is flat, again signifiying that the system is limited by external vibrations in this regime. The red dotted line represents the vibrational noise across all $Q$. (d) Shows the same as (c) for the $\beta$ mode.}
    \label{fig:calibration}
\end{figure*}

\begin{equation}
    U = \frac{\mu_0\mu^2}{64 \pi z^3}(1 + sin^2\beta) + mgz,
    \label{eq:Potential_energy}
\end{equation}

where $g$ is acceleration due to gravity. We find the equilibrium position of the levitated magnet by minimization of this potential. The minimum position is achieved at $z = z_0$ and $\beta = \beta_0 = 0$. $z_0$ is the equilibrium height and is given by

\begin{equation}
    z_0 = \left(\frac{3\mu_0\mu^2}{64 \pi m g} \right)^{1/4}.
\end{equation}

The resonance frequencies are given by $\omega_z = \sqrt{k_z/m}$ and $\omega_{\beta} = \sqrt{k_{\beta}/I}$, where $k_z$ and $k_{\beta}$ are spring constants and $I$ is the moment of inertia. the Spring constants are given by:

\begin{align}
    k_z &= \left[\frac{d^2U}{dz^2}\right]_{(z, \beta) = (z_0, \beta_0)}, \\
     k_{\beta} &= \left[\frac{d^2U}{dz^2}\right]_{(z, \beta) = (z_0, \beta_0)}.
\end{align}

Our levitated magnet is a cylinder, with perpendicular moment of inertia $I = \frac{1}{12}m(d^2 + 3r^2)$. The modal frequencies are given by:

\begin{align}
    \omega_z &= \sqrt{\frac{4g}{z_0}}, \\
    \omega_{\beta} &= \sqrt{\frac{2z_0gm}{3I}}.
    \label{eq:freqs}
\end{align}

If we use eqs.~\eqref{eq:Potential_energy} and \eqref{eq:freqs}, and the parameters detailed in table~\ref{table}, we find values of $\omega_z/2\pi$ = 39.7~Hz and $\omega_{\beta}/2\pi$ = 175.4~Hz. Both of these values are close to the real values (6\% and 2\% difference, respectively), and are shown in Fig.~2~(a) in the main manuscript.

\section*{Appendix D: Calibration of normal modes}

In the main manuscript, the temperatures that are measured are calibrated relative to data taken at a high background gas pressure. Taking the reference data at high pressure means the system is thermal noise limited, and therefore the temperature, and by extension the amplitude, of the modes are known. In order to justify that our system is indeed thermal noise limited at this calibration pressure, data taken at the calibration pressure ($1 \times 10^{-1}$~mbar) is compared to data taken at lower pressures (with higher Q-factors). In Fig.~\ref{fig:calibration}, these measurements can be seen for the $z$ and $\beta$ mode. For a system to be thermal noise limited, the thermal noise mechanism, in this gas the gas particles in the vacuum chamber, must drive and damp the system to be in equilibrium with the thermal bath. As the thermal noise decreases ($Q$ increases), the amplitude should remain at the same level, while still thermally limited. In Fig.~\ref{fig:calibration}~(a) and~(b), we have plotted the normalized RMS amplitude vs $Q$ for the $z$ and $\beta$ mode respectively. As can be seen, the amplitude remains constant while thermally limited up to a $Q\sim10^4$. At higher $Q$ factors, the thermal noise is below that of the excess vibrational noise, which now dominates the motion. To convince ourselves that this vibrational noise is indeed what is driving the system, we plot the ratio $V_{RMS}/\sqrt{Q}$ vs $Q$ in Fig.~\ref{fig:calibration}~(c) and~(d). For a consistent vibrational noise, this ratio, which represents the noise power, should remain constant across the region where the system is limited by external vibrations, and the ratio should increase when the system is thermal limited, as seen for both the $z$ and $\beta$ mode at our calibration pressure. This means that we can use the thermally limited data as a calibration for all data which is not thermally limited. 

\section*{Appendix E: Optimization of SQUID coupling}

\begin{figure}[t!]
    \centering
    \includegraphics[width=0.9\linewidth]{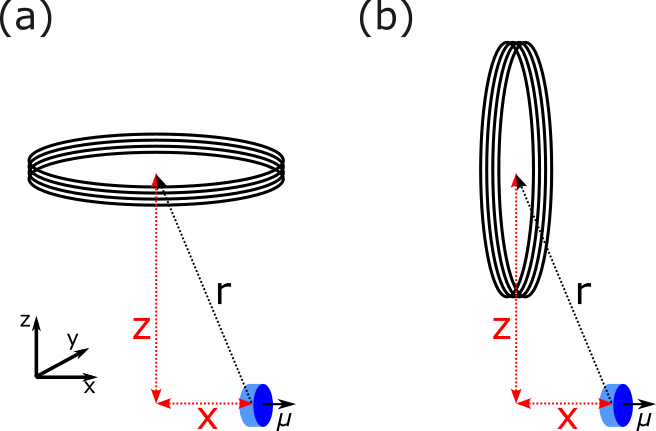}
    \caption{Schematic showing the pick-up coil and magnet geometry. A pick-up coil of $N$ turns is placed in the vicinity of a magnetic dipole $\bm{\mu}$. (a) Shows the geometry currently used for detection. The pick-up coil is placed at a vertical distance $z$ from the magnet, and misaligned by a small amount $x$, to ensure all normal modes couple into the pick-up coil. (b) Shows a proposed coil geometry where the coil is oriented at 90 degrees to the current geometry. Such a geometry will be more strongly coupled to the $z$-mode, while being less coupled to $x$ and $y$ modes.}
    \label{fig:Detection_optimisation}
\end{figure}

In our current experiment, our pick-up coil is not optimized for detecting any particular mode. Therefore, tailoring the geometry and orientation of the pick-up coil to be sensitive to the $z$-mode can substantially increase the magnetomechanical coupling of motion into the pick-up coil, and therefore the SQUID. Here, we consider the magnetic flux coupled into the pick-up coil as the magnet moves in the $z$-direction. The magnetic flux through a loop of area $\bm{A}$ given by the surface integral $\Phi_B = \iint_S \bm{B} \cdot d\bm{A}$. By considering our magnet to be a magnetic dipole $\bm{\mu}$ and our pick-up coil to be made of $N$ circular loops of radius $R$, we have that 

\begin{equation}
    \Phi_B = N \bm{B} \cdot \bm{A},
    \label{eq:PhiB}
\end{equation}

where $\bm{B}$ is the magnetic field produced by the magnet at the pick-up coil and $\bm{A}$ is the vector area of the pick-up coil. The magnetic field produced by a dipole is given by

\begin{equation}
    \bm{B(r)} = \frac{\mu_0}{4\pi r^3}(3\bm{(\mu\cdot \hat{r})\hat{r}-\mu}),
    \label{eq:Dipole}
\end{equation}

where $\bm{r}$ is a position vector with unit vector $\bm{\hat{r}}$. By defining the unit position vector as 

\begin{equation}
    \bm{\hat{r}} = \frac{\bm{r}}{r} = \frac{x\bm{\hat{x}} + z\bm{\hat{z}}}{\sqrt{x^2+z^2}},
\end{equation}

where $x$ and $z$ are the horizontal and vertical distances, as shown in~Fig.~\ref{fig:Detection_optimisation}, we find that the magnetic field is

\begin{equation}
    \bm{B(r)} = \frac{\mu_0}{4\pi}\left(\frac{3\mu x(x\bm{\hat{x}} + z\bm{\hat{z}})}{(x^2+z^2)^2} - \frac{\bm{\mu}}{(x^2+z^2)^{3/2}}\right).
    \label{eq:Dipole}
\end{equation}

By plugging this into eq.~\eqref{eq:PhiB}, we have an expression for the magnetic flux through the pick-up coil, which is

\begin{equation}
    \Phi_B = N\frac{\mu_0}{4\pi}\left(\frac{3\mu x(x\bm{\hat{x}} + z\bm{\hat{z}})}{(x^2+z^2)^2} - \frac{\bm{\mu}}{(x^2+z^2)^{3/2}}\right)\cdot \pi R^2 \bm{\hat{n}},
\end{equation}

where $\pi R^2\bm{\hat{n}}$ is the area vector of the pick-up coil. We consider the case when the area vector direction $\bm{\hat{n}}$ is perpendicular to $\bm{\mu}$, as in our current experiment (Fig.~\ref{fig:Detection_optimisation}~(a)), and the case when the area vector is parallel to $\bm{\mu}$ (Fig.~\ref{fig:Detection_optimisation}~(b)). In these orientations, the magnetic flux through the pick-up loop is given by

\begin{align}
    \Phi_{B\perp} &= \frac{3N\mu_0R^2\mu xz}{4(x^2+z^2)^2},\\ \Phi_{B\parallel} &= \frac{N\mu_0R^2\mu}{4}\left(\frac{3x^2}{(x^2+z^2)^2}-\frac{1}{(x^2+z^2)^{5/2}}\right).
    \label{eq:Phi_pp}
\end{align}

In order to compare how the couplings of either orientation compare for the $z$-mode, we differentiate $\Phi_{B\perp}$ and $\Phi_{B\parallel}$ with respect to $z$, resulting in

\begin{align}
    \frac{\partial \Phi_{B\perp}}{\partial z} &= \frac{3N\mu_0R^2\mu x(x^2-3z^2)}{4(x^2+z^2)^3}, 
    \label{eq:dPhi_dz_perp}\\
    \frac{\partial \Phi_{B\parallel}}{\partial z} &= \frac{N\mu_0R^2\mu}{4}\left(\frac{3z}{(x^2+z^2)^{5/2}}-\frac{12x^2z}{(x^2+z^2)^3}\right).
    \label{eq:dPhi_dz_parallel}
\end{align}

Experimentally, in the current setup, we have values of $N=15$, $x=0.3$~mm, $z=2.5$~mm, $R=1.0$~mm and $\mu = \frac{1}{\mu_0}B_rV = 1.4 \times 10^{-6} \ \text{A}\cdot \text{m}^2$, where $B_r$ is the residual flux density of the magnet, and $V$ the volume of the magnet. Plugging these numbers into eq.~\eqref{eq:dPhi_dz_perp} and ~\eqref{eq:dPhi_dz_parallel} we find that $\lvert\frac{\partial \Phi_{B\perp}}{\partial z}\rvert = 4.24 \times 10^{-10}$~Wb/m and $\lvert\frac{\partial \Phi_{B\parallel}}{\partial z}\rvert = 4.77 \times 10^{-7}$~Wb/m. The ratio of these couplings is $\lvert\frac{\partial \Phi_{B\parallel}}{\partial z}\rvert/\lvert\frac{\partial \Phi_{B\perp}}{\partial z}\rvert = 1100$, meaning that rotating the pick-up coil by 90 degrees would result in a detection noise of $S_{x_{d}}^{1/2} = 1.3 \times 10^{-16}$~m/$\sqrt{\text{Hz}}$. We find that this translates to a minimum phonon number $N_{\text{min}}^z \sim 2$, or 5~nK in temperature (see eq.~(5) in the main manuscript). Given the $1/z^4$ dependence in eq.~\eqref{eq:dPhi_dz_parallel}, reducing $z$ from $z = 2.5$~mm to $z = 2.0$~mm would result in a magnetomechanical coupling which is strong enough to reach the quantum ground state.

In order to reach the ground state, the force noise must be dominated by SQUID backaction, and the detector itself be quantum limited, such that the energy resolution of the SQUID $\epsilon \approx \hbar$~\cite{Braginsky1995, Clerk2010, Caves1982}. The backaction from SQUIDs which are not quantum limited, would set a limit on cooling such that $N_{\text{min}}>1$. SQUIDs approaching such a the quantum limit have been achieved~\cite{Wakai1988, Awschalom1988}. With state-of-the-art SQUID detection, it is feasible to achieve the necessary conditions to cool towards the quantum ground state.

\section*{Appendix F: Experimental Parameters}

This table lists the physical parameters of the magnet, as well as measured and predicted experimental parameters of the $z$ and $\beta$ mechanical mode.

\begin{table*}
\begin{tabularx}{\textwidth}{|l|X|l|l|l|}
\hline
 & \textbf{Parameter} & \textbf{Symbol} & \textbf{Value} & \textbf{Unit} \\
 \hline 
 
 \textbf{Properties of magnet} & Mass & m & 23  & $\mu$g \\ 
 & Radius & r & 100  & $\mu$m \\ 
 & Thickness & d & 100  & $\mu$m \\
 & Density & $\rho$ & 7430  & $kg/m^3$ \\
 & Perpendicular moment of inertia & I & $7.8 \times 10^{-17}$ & $kg\cdot m^2$ \\
 & Magnetization & M & $4.4 \times 10^5$  & A/m \\
 \hline
 
 \textbf{z-mode parameters} & Frequency & $\omega_z/2\pi$ & 42.4  & Hz \\ 
 & Quality factor & $Q_z$ & $1.0 \times 10^7$  & - \\
 & Equilibrium effective temperature & $T_{\text{eq}}^{z}$ & 3400  & K \\
 & Equilibrium amplitude & $z_{\text{eq}}$ & 7.6 & nm \\
 & Force noise & $S_F^{1/2}$ & $3.4 \times 10^{-16}$ & N/$\sqrt{\text{Hz}}$ \\
 & Detection noise & $S_{x_{d}}^{1/2}$ & $2.1 \times 10^{-11}$ & m/$\sqrt{\text{Hz}}$ \\
 & Voltage to displacement conversion factor & $C_z$ & $1.76 \times 10^{6}$ & V/m \\
 & Feedback cooled effective temperature & $T_{\text{FB}}^{z}$ & $\sim 7$  & K \\
 & Feedback cooled amplitude & $z_{\text{FB}}$ & $\sim340$ & pm \\
 & Minimum possible feedback cooled temperature & $T_{\text{min}}^{z}$ & 70  & mK \\ 
 & Minimum possible feedback cooled amplitude & $z_{\text{min}}$ & 34 & pm \\
 & SQUID noise floor at 42.4~Hz & $S_{\Phi_0}^{1/2}$ & $\sim 0.6$ & $\mu \Phi_0/\sqrt{\text{Hz}}$ \\ 
  
 \hline
 \textbf{$\boldsymbol{\beta}$-mode parameters} & Frequency & $\omega_z/2\pi$ & 178.8  & Hz \\ 
 & Quality factor & $Q_{\beta}$ & $2.1 \times 10^6$  & - \\
 & Equilibrium effective temperature & $T_{\text{eq}}^{\beta}$ & 97 & K \\
 & Equilibrium amplitude & $\beta_{\text{eq}}$ & 5.2 & $\mu$rad \\
 & Torque noise & $S_{\tau}^{1/2}$ & $1.5 \times 10^{-20}$ & N${\cdot}$m/$\sqrt{\text{Hz}}$ \\
 & Detection noise & $S_{\theta_{d}}^{1/2}$ & $3.0 \times 10^{-7}$ & rad/$\sqrt{\text{Hz}}$ \\
 & Voltage to angle conversion factor & $C_{\beta}$ & 123.2 & V/rad \\
 & Feedback cooled effective temperature & $T_{\text{FB}}^{\beta}$ & $830\pm200$  & mK \\
 & Feedback cooled amplitude & $\beta_{\text{FB}}$ & $480\pm100$ & nrad \\
 & Minimum possible feedback cooled temperature & $T_{\text{min}}^{\beta}$ & 180  & mK \\
 & Minimum possible feedback cooled amplitude & $\beta_{\text{FB}}$ & 220 & nrad \\
 & SQUID noise floor at 178.8~Hz & $S_{\Phi_0}^{1/2}$ & $\sim 0.5$ & $\mu \Phi_0/\sqrt{\text{Hz}}$ \\
 \hline
 
 \textbf{z-mode expected future parameters} & Force noise at thermal equilibrium & $S_F^{1/2}$ & $3.7 \times 10^{-18}$ & N/$\sqrt{\text{Hz}}$ \\
 & Detection noise at SQUID noise floor & $S_{x_{d}}^{1/2}$ & $1.4 \times 10^{-13}$ & m/$\sqrt{\text{Hz}}$ \\
 & Minimum possible feedback temperature at thermal noise and SQUID noise limit & $T_{\text{min}}^z$ & $\sim$ 5  & $\mu$K \\
 & Minimum possible feedback cooled amplitude at thermal noise and SQUID noise limit & $z_{\text{min}}$ & 290 & fm \\
 & Minimum phonon number & $N_{min}^{z}$ & $\sim$ 2500  & -\\
 \hline
 
 \textbf{$\boldsymbol{\beta}$-mode expected future parameters} & Torque noise at thermal equilibrium & $S_{\tau}^{1/2}$ & $9.8 \times 10^{-22}$ & Nm/$\sqrt{\text{Hz}}$ \\
 & Detection noise at SQUID noise floor & $S_{\theta_{d}}^{1/2}$ & $2.1 \times 10^{-9}$ & rad/$\sqrt{\text{Hz}}$ \\
 & Minimum possible feedback temperature at thermal noise and SQUID noise limit & $T_{\text{min}}^{\beta}$ & $\sim$ 80  & $\mu$K\\
 & Minimum possible feedback cooled amplitude at thermal noise and SQUID noise limit & $\beta_{\text{FB}}$ & 4.8 & nrad \\
 & Minimum phonon number & $N_{min}^{\beta}$ & $\sim$ 9400  & -\\

\hline
 
\end{tabularx}
\caption{Table showing the relevant experimental parameters of the magnet and normal modes of interest. In the top row basic parameters of the magnet itself are listed, followed by the measured parameters of the z-mode and $\beta$-mode respectively. Finally, predicted experimental values for the z and $\beta$-mode are listed. These predictions are based on introducing sufficient vibration isolation to reach the thermal noise limit, and by measuring to the full capability of the SQUID used for readout. Currently the detection noise is limited by the data acquisition, rather than SQUID noise.}
\label{table}
\end{table*}

\clearpage

\bibliography{references.bib}

\appendix



\end{document}